\documentclass[twocolumn,floatfix,eqsecnum,aps]{revtex4-2}
\usepackage{graphicx}
\usepackage{amsmath}
\usepackage{amssymb}
\usepackage{bm}
\usepackage{hyperref}

\usepackage{color}

\begin{document}

\title{Quantum circuit decomposition of the tangent-fermion Dirac operator}
\author{C. W. J. Beenakker}
\affiliation{Instituut-Lorentz, Universiteit Leiden, P.O. Box 9506, 2300 RA Leiden, The Netherlands}

\date{June 2026}

\begin{abstract}
The Dirac operator on a lattice cannot be both local and free of fermion doubling, at least not without breaking fundamental symmetries. Non-local, symmetry-preserving discretizations that avoid doubling have a quantum circuit representation as a linear-combination-of-unitaries (LCU) in which both the number of terms and their norm (the subnormalization factor) grow with the lattice size, compromising the efficiency of a quantum algorithm. We show that the tangent-fermion discretization escapes this obstruction when the Dirac equation is written as a generalized eigenvalue problem with a local operator pencil: Each member of the pencil has an exact LCU, with term count that is independent of lattice size and with subnormalization factor of order unity, on a par with elliptic operators. This provides an efficient block-encoding primitive for Dirac spectra and Green functions without fermion doubling.
\end{abstract}
\maketitle

\section{Introduction}

While products of unitary operators can be implemented deterministically on a quantum computer, sums of unitaries require a probabilistic implementation, conditioned on a measurement outcome \cite{Chi12,Chi17,Low19,Gil19,Ral20}. 

The computational efficiency of the LCU (linear combination of unitaries) decomposition of a non-unitary operator $A$,
\begin{equation}
A=\sum_{k=1}^L c_k U_k,\;\;c_k>0,\;\;U_k^{\vphantom{\dagger}}U_k^\dagger=\openone,\label{LCUA}
\end{equation}
depends not only on the number $L$ of unitaries needed, but also on the so-called subnormalization factor \cite{note1}
\begin{equation}
\lambda=\|c\|_1=\sum_{k=1}^L |c_k|,
\end{equation}
which governs the success probability $P_{\rm success}\propto 1/\lambda^2$. A larger $\lambda$ means the desired quantum state $A\Psi$ is generated less frequently, potentially compromising the speed-up of the quantum algorithm.

A first step towards a quantum solver of a partial differential equation is to represent the discretized differential operator by an LCU with few unitaries and small subnormalization factor. Here we work this out for the Dirac operator in the tangent fermion discretization \cite{Sta82,Bee23}. We compare with the LCU decomposition of elliptic operators \cite{Chi21,Sat21,Li23,Kha25,Hog26} and of the Dirac operator \cite{Gha26} in the \textsc{slac} fermion discretization \cite{Dre76}.

Since the Dirac operator is a first order differential, its discretization faces the fermion-doubling obstruction \cite{Nie81}: The finite difference $f'(x)\mapsto (2a)^{-1}[f(x+a)-f(x-a)]$ has Fourier representation $ik F(k)\mapsto (i/a)F(k)\sin ak$, which vanishes not only at $k=0$, as it should, but also at the boundary $k=\pi/a$ of the Brillouin zone. The spurious zero produces a second species of low-energy excitations (hence the name ``fermion doubling''), which breaks the topological protection of a massless Dirac fermion \cite{Tong}.

Fermion doubling can be avoided, without breaking the fundamental chiral symmetry of massless Dirac fermions, by a \textit{nonlocal} finite difference scheme that couples distant lattice sites. Two classic examples are the \textsc{slac} fermion discretization \cite{Dre76}, with a coupling that decays as 1/distance, and the tangent fermion discretization \cite{Sta82}, with an all-to-all coupling that does not decay at all. The corresponding LCU decomposition on an $N=2^n$-site lattice contains $L={\cal O}(N)$ unitaries for both \textsc{slac} and tangent fermions, with subnormalization factors $\lambda$ of order $O(n)$ and $O(2^{n})$, respectively. Used directly, the first demands the sophisticated dense-state preparation machinery of Ref.~\onlinecite{Gha26}; the second is unusable.

The purpose of this note is to point out an indirect route for the tangent fermion LCU that can bring both $L$ and $\lambda$ down to $O(1)$, on a par with elliptic operators. The indirect route replaces the discretized Dirac equation $H_{\rm Dirac}\Psi=E\Psi$ by a generalized eigenproblem ${\cal H}\Psi=E{\cal P}\Psi$, with local Hermitian operators ${\cal H}$ (a nearest-neighbour difference)  and ${\cal P}$ (a nearest-neighbour average) \cite{Pac21}. Each admits an exact two- or three-term LCU whose unitaries are a single (anti)periodic shift gate. We outline two possible applications of this representation, to quantum calculations of the spectrum and Green function of the Dirac operator.

\section{Discretization of the Dirac equation}

We discretize the one-dimensional (1D) Dirac equation
\begin{equation}
\begin{split}
&H_{\rm Dirac}\Psi=E\Psi,\\
&H_{\rm Dirac}=vp_x\sigma_x=-i\hbar v\begin{pmatrix}
0&d/dx\\
d/dx&0
\end{pmatrix},
\end{split}
\end{equation}
on a 1D lattice (lattice constant $a$). The spinor $\Psi$ has two components, acted upon by the Pauli matrix $\sigma_x$. The 1D case is considered for ease of notation, generalizations to higher dimensions are straightforward and will be indicated later on. In what follows we set $\hbar v$ and $a$ equal to unity.

We compare three alternative discretizations of the differential operator $d/d x$, one local scheme and two nonlocal schemes:
\begin{align}
\partial^{\rm sine} ={}& \tfrac{1}{2}(T-T^{-1}),\;\;\text{Fourier symbol } i\sin k,
\label{eq:central}\\
\partial^{\rm sawtooth} ={}&  \ln T = \sum_{j=1}^{\infty}\frac{(-1)^{j}}{j}\,(T^{-j}-T^{j}), \nonumber\\
&\text{Fourier symbol } \,\ln e^{ik} ,
\label{eq:slac}\\
\partial^{\rm tangent} ={}& 2\,\frac{T-1}{T+1} = 2\sum_{j=1}^{\infty}(-1)^{j}\,(T^{-j}-T^{j}),\nonumber\\
& \text{Fourier symbol } 2i\tan( k/2).
\label{eq:tangent}
\end{align}
Here $T=e^{d/d x}$ is the translation operator, $(T f)(x)=f(x+1)$, with Fourier symbol $e^{ik}$. 

The central finite difference $\partial^{\rm sine}$ is a local discretization, only neighboring sites are coupled. Its Fourier symbol $\sin ak$ has a spurious zero (doubler) at the boundary $k=\pi$ of the Brillouin zone. The nonlocal discretization $\partial^{\rm sawtooth}$ implements the \textsc{slac} derivative \cite{Dre76}, with a linear symbol up to the zone boundary, where it jumps in a sawtooth fashion. The tangent symbol of the nonlocal Stacey derivative \cite{Sta82} $\partial^{\rm tangent}$, instead, is linear only near $k=0$ and has a pole at the zone boundary.

The \textit{nonlocal} tangent discretization of the Dirac equation can equivalently be written in the form of a \textit{local} generalized eigenproblem \cite{Pac21},
\begin{equation}
(\sigma_x K-EB)\Psi = 0,
\end{equation}
with operator pencil $(\sigma_x K,B)$ given by
\begin{equation}
\begin{split}
& K=\tfrac{1}{2}i(T^{-1}-T),\;\;\text{symbol } \sin k, \\
&B=\tfrac{1}{4}(1+T^{-1})(1+T),\;\;\text{symbol } \tfrac{1}{2}(1+\cos k).
\end{split}
\end{equation}
These are both sparse Hermitian matrices, only neighboring sites are coupled.

\section{LCU decomposition}

\subsection{1D case}

We consider a finite lattice of $N=2^n$ sites encoded into $n$ qubits: site $x=j$ with binary string $j$ is represented by the state $|j\rangle$ in which the $p$-th digit of $j$ gives the state of the $p$-th qubit. One more qubit encodes the spin degree of freedom.

We impose antiperiodic boundary conditions,
\begin{equation}
|j+N\rangle=-|j\rangle,\;\; j\in\{0,1,\ldots N-1\}.
\end{equation}
The antiperiodic translation operator
\begin{equation}
\tilde{T}|j\rangle=\begin{cases}
|j+1\rangle&\text{if}\;\;j\in\{0,1,\ldots N-2\},\\
-|0\rangle&\text{if}\;\;j=N-1,
\end{cases}
\end{equation}
is implemented as the product $\tilde{T}=S C$ of the cyclic shift $S$ and the multi-controlled-Z operator $C$,
\begin{equation}
\begin{split}
&S|j\rangle=|j+1\bmod{N}\rangle,\;\;C|j\rangle=(-1)^{\delta_{j,N-1}}|j\rangle,\\
&j\in\{0,1,2,\ldots N-1\}.
\end{split}
\end{equation}
The ${\rm U}(N)$ matrix representation of $\tilde{T}$ is
\begin{equation}
\tilde{T}=S C =
\begin{pmatrix}
0      &        &        & -1 \\
1      & 0      &        &   \\
& \ddots & \ddots &   \\
&        & 1      & 0
\end{pmatrix}.
\end{equation}
The $S$ and $C$ operations need $O(n)$ elementary gates ($\textsc{cnot}$ or single-qubit unitaries), plus one ancilla qubit for the carry logic \cite{Gid15,Bar95}.

The anti-periodic shift operator is unitary, $\tilde{T}^{-1}=\tilde{T}^\dagger$. Since $\tilde{T}^N=-\openone$ the eigenvalues $e^{iq_{j}}$ of $\tilde{T}$ are on the half-integer momentum grid, $q_{j}=(2j+1)\pi/N$, containing neither $q=0$ nor $q=\pi$. This ensures that both operators
\begin{equation}
 \tilde{K}=\tfrac{1}{2}i\tilde{T}^\dagger-\tfrac{1}{2}i\tilde{T}^{\vphantom{\dagger}},\;\;
\tilde{B}=\tfrac{1}{2}\openone+\tfrac{1}{4}\tilde{T}^\dagger+\tfrac{1}{4}\tilde{T}^{\vphantom{\dagger}}\label{tildeKB}
\end{equation}
are invertible. 

Eq.\ \eqref{tildeKB} is an LCU decomposition of the $(\sigma_x\tilde{K},\tilde{B})$ pencil with unit subnormalization factor $\lambda_{\rm tangent}=1$, the same as for the sine discretization \eqref{eq:central}, $\lambda_{\rm sine}=1$. The local generalized eigenvalue (pencil) representation is essential here: The nonlocal tangent fermion derivative operator \eqref{eq:tangent} has on the $N$-site lattice an exponentially large subnormalization factor $2(N-1)=O(2^n)$. The \textsc{slac} derivative \eqref{eq:slac} does not have a local pencil representation, its subnormalization factor is $\lambda_{\rm sawtooth}=O(n)$.

In Table \ref{tab:counts} we compare the figures of merit for the LCU decompositions of the three discretizations.

\begin{table*}[t]
\centering
\small
\begin{tabular}{lcccc}
\hline
 & \# of terms $L$ & $\ell_{1}$ norm $\lambda$ & \# of gates/term & no doublers?\\
\hline
$\partial^{\rm sine}$ & $2$ & $1$ & $O(n)$ & $\times$\\
$\partial^{\rm sawtooth}$& $O(2^{n})$ & $O(n)$ & $O(n)$ & $\checkmark$\\
$\partial^{\rm tangent}$  & $O(2^{n})$ & $O(2^n)$ & $O(n)$ & $\checkmark$\\
pencil $(\sigma_x K,B)$ & $(2,3)$ & $(1,1)$ & $O(n)$ & $\checkmark$\\
\hline
\end{tabular}
\caption{LCU data for 1D lattice momentum operators on $N=2^{n}$ sites. The pencil row lists the operators in the generalized eigenproblem $(\sigma_x K-EB)\Psi=0$ for the tangent discretization.}
\label{tab:counts}
\end{table*}

\subsection{Higher dimensions}

In $d$ dimensions the Dirac equation reads
\begin{equation}
-i\sum_{\alpha=1}^d \Gamma_\alpha \frac{\partial}{\partial x_\alpha}\Psi=E\Psi,
\end{equation}
with $\Gamma_1,\Gamma_2,\ldots \Gamma_d$ a set of mutually anti-commuting unitary Hermitian matrices (Clifford algebra). The spinor $\Psi$ now has length $2^{\lfloor d/2\rfloor}$, encoded into $\lfloor d/2\rfloor$ qubits.

The $N^d=2^{nd}$ lattice sites on a hypercubic lattice are encoded into $d$ registers of $n$ qubits each. Antiperiodic boundary conditions are applied in each $x_\alpha$ direction. Translation operators $\tilde{T}_\alpha$ act only on register $\alpha$.

The nonlocal tangent discretization of the differential operator still has a local generalized eigenproblem representation,
\begin{equation}
({\cal H}-E{\cal P})\Psi=0,\label{HPgeneigen}
\end{equation}
with operator pencil $({\cal H},{\cal P})$ now given by \cite{Pac21}
\begin{subequations}
\begin{align}
&{\cal H}=\sum_{\alpha=1}^d\Gamma_\alpha \tilde{K}_\alpha\prod_{\beta\neq\alpha}^d \tilde{B}_\beta,\;\;{\cal P}=\prod_{\alpha=1}^d\tilde{B}_\alpha,\\
& \tilde{K}_\alpha=\tfrac{1}{2}i(\tilde{T}_\alpha^{\dagger}-\tilde{T}_\alpha^{\vphantom{\dagger}}),\;\;\tilde{B}_\alpha=\tfrac{1}{4}(\openone+\tilde{T}_\alpha^{\dagger})(\openone+\tilde{T}_\alpha^{\vphantom{\dagger}}).
\end{align}
\end{subequations}
The LCU subnormalization factors are\begin{equation}
\lambda_{\rm tangent}({\cal H})=d,\;\;
\lambda_{\rm tangent}({\cal P})=1,
\end{equation}
since $\lambda(\prod_\alpha A_\alpha)=\prod_\alpha\lambda(A_\alpha)$. The number $L$ of terms in the LCU decomposition is given by
\begin{equation}
L_{\rm tangent}({\cal H})=2d3^{d-1},\;\;L_{\rm tangent}({\cal P})=3^d.
\end{equation}

\subsection{Mass and scalar potentials}

To make the Dirac fermions massive one needs $d+1$ anticommuting unitary Hermitian matrices, which then have dimension $2^{\lfloor (d+1)/2\rfloor}$. For example, in 2D the mass term in the Dirac equation is $\mu\sigma_z$, with Pauli matrix $\sigma_z$ anticommuting with $\sigma_x$ and $\sigma_y$. More generally, we denote the mass term by $\mu\Gamma_{d+1}$. In any dimension, an electrical potential $V$ enters as a scalar, multiplying the unit matrix $\Gamma_0$. Both may be position dependent fields $\mu(\bm{r}),V(\bm{r})$.

The generalized eigenproblem $({\cal H}-E{\cal P})\Psi=0$ is modified into \cite{Pac21}
\begin{equation}
\begin{split}
&{\cal H}=\sum_{\alpha=1}^d\Gamma_\alpha \tilde{K}_\alpha\prod_{\beta\neq\alpha}^d \tilde{B}_\beta+\Phi^\dagger(\mu\Gamma_{d+1}+V\Gamma_0)\Phi,\\
&{\cal P}=\Phi^\dagger\Phi,\;\;\Phi=\prod_{\alpha=1}^d\Phi_\alpha,\;\; \Phi_\alpha=\tfrac{1}{2}(\openone+\tilde{T}_\alpha).
\end{split}\label{HPPhi}
\end{equation}
The operators ${\cal H}$ and ${\cal P}$ remain local and Hermitian, but ${\cal H}$ no longer commutes with ${\cal P}$ if translational invariance is broken by $\mu$ or $V$.

The LCU decomposition of ${\cal H}$ now needs a representation of the fields $\mu$ and $V$ as sums of unitaries. This can be realized efficiently if the inhomogeneities are spatially localized, since a projector $|\bm{r}\rangle\langle \bm{r}|$ has the 2-term LCU decomposition
\begin{equation}
|\bm{r}\rangle\langle \bm{r}|=\tfrac{1}{2}\openone+\tfrac{1}{2}U_{\bm r},\;\;U_{\bm r}=2|\bm{r}\rangle\langle\bm{r}|-\openone,\;\;U_{\bm r}^{\vphantom{\dagger}}U_{\bm r}^\dagger=\openone.
\end{equation}
Few-mode Fourier series or low-degree polynomial fields also have efficient LCU decompositions, while unstructured fields are problematic \cite{Kuk25}. 

\section{Applications of the block encoded Dirac operator}

The LCU decomposition \eqref{LCUA} of a linear operator $A$ enables its block encoding into a unitary ${\cal U}$ acting on system plus ancilla qubits \cite{Low19},
\begin{equation}
{\cal U}=\begin{pmatrix}
A/\lambda&\ast\\
\ast&\ast
\end{pmatrix},
\end{equation}
where $\ast$ can be any block matrix. This is a primitive for a variety of quantum algorithms \cite{Cha19}. We discuss two applications of the LCU decomposition of the Dirac operator $H_{\rm Dirac}$, in the tangent fermion pencil representation $({\cal H},{\cal P})$. To allow for the case of broken translational invariance, we will not assume that ${\cal H}$ and ${\cal P}$ commute.

\subsection{Spectral analysis}
\label{sec:spectral}

The first application is to the solution of generalized eigenvalue problems \cite{Par20,Lia22,Sha22,Raj26}. We seek the spectrum of the tangent fermion Dirac equation, $({\cal H}-E{\cal P})\Psi=0$, focusing on eigenvalues near the Dirac point $E=0$. Because the Dirac point sits in the middle of the band (for free Dirac fermions the spectrum is $\pm E$ symmetric), we cannot use ground-state solvers.

Variational minimization of the Rayleigh quotient $\langle\Psi|{\cal H}|\Psi\rangle/\langle\Psi|{\cal P}|\Psi\rangle$ is cheap per iteration --- only the few-term LCU decompositions of ${\cal H}$ and ${\cal P}$ are needed to evaluate it, for a trial state $\Psi$ prepared by a parameterized quantum circuit --- but it converges to the band edges at large $|E|$, not to the interior. Reaching energies near $E=0$ variationally requires the folded quotient $\langle\Psi|{\cal H}{\cal P}^{-1}{\cal H}|\Psi\rangle/\langle\Psi|{\cal P}|\Psi\rangle$, whose numerator no longer has an efficient LCU decomposition because of the ${\cal P}^{-1}$. Indeed, the smallest eigenvalue of ${\cal P}$ on the antiperiodic grid is $\sin^2(\pi/2N)=O(N^{-2})$, at the grid point closest to the Brillouin zone boundary, which implies a large subnormalization factor $\lambda({\cal P}^{-1})\geq\|{\cal P}^{-1}\|=O(N^2)$.

An approach that does not require inversion of ${\cal P}$ is the ``quantum landscape scanning'' method of Ref.\ \onlinecite{Raj26}. The smallest singular value $\sigma_{\rm min}(E)$ of the matrix $M(E)={\cal H}-E{\cal P}$ is determined as a function of the real energy $E$, scanned on a grid of $N_E$ points around $E=0$. Local minima in this ``landscape'' align with generalized eigenvalues.

Each trial energy needs only a block encoding of $M(E)$, assembled additively from those of ${\cal H}$ and ${\cal P}$. The smallest singular value is obtained by phase estimation, with complexity $O(\sqrt{NN_E}/\varepsilon)$ for precision $\varepsilon$. This is quadratically better in $N$ and $N_E$ than a classical scan, at the price of a $1/\varepsilon$ rather than $\log(1/\varepsilon)$ precision dependence \cite{Raj26}.

\subsection{Green function}
\label{sec:green}

The Green function (resolvent) of the Dirac operator is given by
\begin{equation}
G(z)=(H_{\rm Dirac}-z)^{-1}=\Phi({\cal H}-z{\cal P})^{-1}\Phi^\dagger,
\end{equation}
with ${\cal P}=\Phi^\dagger\Phi$ from Eq.\ \eqref{HPPhi}. This quantity gives access to the local density of states and to the scattering amplitudes that determine transport properties. Its matrix elements are
\begin{equation}
G_{ab}(z)=\langle a|G(z)|b\rangle=\langle \bar{a}|\Psi_b(z)\rangle,\label{Gab}
\end{equation}
with $|\bar{a}\rangle=\Phi^\dagger|a\rangle$, $|\bar{b}\rangle=\Phi^\dagger|b\rangle$, and $\Psi_b(z)$ determined by the inhomogeneous equation
\begin{equation}
({\cal H}-z{\cal P})|\Psi_b(z)\rangle=| \bar{b}\rangle.\label{Psibeq}
\end{equation}

The efficient block encoding of ${\cal H}$ and ${\cal P}$ allows for an efficient solution using established quantum linear system solvers \cite{Har09,Bra23,Tra23,Mor24}, with one modification: To avoid working with a non-Hermitian matrix when $z$ is complex, one may double the dimension and replace Eq.\ \eqref{Psibeq} by
\begin{equation}
\begin{pmatrix}
0&{\cal H}-z{\cal P}\\
{\cal H}-z^\ast{\cal P}&0
\end{pmatrix}
\begin{pmatrix}
0\\
|\Psi_b(z)\rangle
\end{pmatrix}=\begin{pmatrix}
| \bar{b}\rangle\\
0
\end{pmatrix}.
\end{equation}

Once the state $|\Psi_b(z)\rangle$ has been prepared by the quantum solver, a matrix element \eqref{Gab} is read out as the overlap $\langle\bar a|\Psi_b(z)\rangle$ by a Hadamard test; the algorithm thus returns selected matrix elements or expectation values, not the full resolvent matrix. Classically, this requires solving a sparse linear system at a cost that grows with $N$. The quantum solver scales polylogarithmically in $N$, with the important proviso that the speed-up is controlled by the condition number of ${\cal H}-z{\cal P}$ and will require a suitable preconditioner (in particular when $\operatorname{Im} z$ is small).

\section{Conclusion}

We have shown that fermion doubling, which forces any local lattice Dirac operator with chiral symmetry to have a spurious zero-mode \cite{Nie81}, need not degrade the quantum-circuit complexity of the Dirac operator. The tangent-fermion discretization \cite{Sta82} is nonlocal, and a direct block encoding of its all-to-all coupling carries an exponentially large subnormalization factor. We work around this obstruction by using the factorization of the discretized Dirac operator into a pair of \emph{local} operators \cite{Pac21}, producing the Hermitian pencil $({\cal H},{\cal P})$. Each member has an exact LCU with $O(1)$ terms and $O(1)$ subnormalization factor, built from a single (anti)periodic shift gate.

The two applications outlined here, spectra by landscape scanning \cite{Raj26} and Green functions by quantum linear-system solvers \cite{Har09}, open a route to quantum simulation of topological materials with an unpaired Dirac cone, required for the topological protection \cite{Bee23}. What remains to be done for the tangent discretization is an end-to-end resource estimate with suitable preconditioners, along the lines of a recent study \cite{Gha26} for the \textsc{slac} discretization \cite{Dre76}.

All of this is for single-particle simulations (first quantization). It would be of interest to lift the block encoding of the tangent fermion Dirac operator to the multi-particle Fock space (second quantization), where one could study the Luttinger liquid physics relevant for quantum spin Hall insulators \cite{Wu06}. For that problem it has been shown that tangent fermions remain massless in the presence of Hubbard interactions \cite{Zak24}, while \textsc{slac} fermions acquire a spurious mass \cite{Wan23}.

\acknowledgments

I used several AI models (Claude Fable/Opus, GPT-5) as an interactive tool to familiarize myself with the block encoding literature. My research is supported by the Netherlands Organisation for Scientific Research (NWO/OCW), as part of Quantum Limits (project number {\sc summit}.1.1016).


\begin{thebibliography}{99}
\bibitem{Chi12} A. M. Childs and N. Wiebe, \textit{Hamiltonian simulation using linear combinations of unitary operations}, Quantum Inf. Comput. \textbf{12}, 901 (2012).
\bibitem{Chi17} A. M. Childs, R. Kothari, and R. D. Somma, \textit{Quantum algorithm for systems of linear equations with exponentially improved dependence on precision}, SIAM J. Comput. \textbf{46}, 1920 (2017).
\bibitem{Low19} G. H. Low and I. L. Chuang, \textit{Hamiltonian simulation by qubitization}, Quantum \textbf{3}, 163 (2019).
\bibitem{Gil19} A. Gily\'en, Y. Su, G. H. Low, and N. Wiebe, \textit{Quantum singular value transformation and beyond: exponential improvements for quantum matrix arithmetics}, Proc. 51st ACM STOC, 193 (2019).
\bibitem{Ral20} P. Rall, \textit{Quantum algorithms for estimating physical quantities using block-encodings}, Phys. Rev. A \textbf{102}, 022408 (2020).
\bibitem{note1} The name ``subnormalization factor'' for the $\ell_1$ norm of the LCU coefficients in Eq.\ \eqref{LCUA}  refers to the embedding (``block encoding'') of $A/\lambda$ into the upper-left block of a larger unitary matrix $U$. The scaling, or subnormalization, of $A$ by a factor $\lambda$ ensures that its spectral norm $\|A\|$ (the maximal singular value) remains below unity, as required by unitarity of $U$. The operation $\Psi\mapsto \lambda^{-1}A\Psi$ is realized with probability $P_{\rm success}=\lambda^{-2}\|A\Psi\|^2$.
\bibitem{Sta82} R. Stacey, \textit{Eliminating lattice fermion doubling}, Phys. Rev. D \textbf{26}, 468 (1982).
\bibitem{Bee23} C. W. J. Beenakker, A. Don\'{i}s Vela, G. Lemut, M. J. Pacholski, and J. Tworzyd{\l}o, \textit{Tangent fermions: Dirac or Majorana fermions on a lattice without fermion doubling}, Annalen der Physik \textbf{535}, 2300081 (2023).
\bibitem{Chi21} A. M. Childs, J.-P. Liu, and A. Ostrander, \textit{High-precision quantum algorithms for partial differential equations}, Quantum \textbf{5}, 574 (2021).
\bibitem{Sat21} Y. Sato, R. Kondo, S. Koide, H. Takamatsu, and N. Imoto, \textit{Variational quantum algorithm based on the minimum potential energy for solving the Poisson equation}, Phys. Rev. A \textbf{104}, 052409 (2021).
\bibitem{Li23} Haoya Li, Hongkang Ni, and Lexing Ying, \textit{On efficient quantum block encoding of pseudo-differential operators}, Quantum \textbf{7}, 1031 (2023).
\bibitem{Kha25} T. Kharazi, A. M. Alkadri, J.-P. Liu, K. K. Mandadapu, and K. B. Whaley, \textit{Explicit block encodings of boundary value problems for many-body elliptic operators}, Quantum \textbf{9}, 1764 (2025).
\bibitem{Hog26} T. Hogancamp, R. Demirdjian, and D. Gunlycke, \textit{A linear combination of unitaries decomposition for the Laplace operator}, arXiv:2601.06370.
\bibitem{Gha26} R. M. Gharat, G. Muraleedharan, D. W. Berry, and G. K. Brennen, \textit{Quantum algorithm for solving differential equations using SLAC derivatives}, arXiv:2605.04861.
\bibitem{Dre76} S. D. Drell, M. Weinstein, and S. Yankielowicz,
\textit{Strong-coupling field theories. II. Fermions and gauge fields on a lattice}, Phys. Rev. D \textbf{14}, 1627 (1976).
\bibitem{Nie81} H. B. Nielsen and M. Ninomiya, \textit{A no-go theorem for regularizing chiral fermions}, Phys. Lett. B \textbf{105}, 219 (1981).
\bibitem{Tong} An overview of methods to avoid fermion doubling in lattice gauge theory can be found in chapter 4 of David Tong's lecture notes: \url{https://www.damtp.cam.ac.uk/user/tong/gaugetheory.html}.
\bibitem{Pac21} M. J. Pacholski, G. Lemut, J. Tworzyd{\l}o, and C. W. J. Beenakker, \textit{Generalized eigenproblem without fermion doubling for Dirac fermions on a lattice}, SciPost Phys. \textbf{11}, 105 (2021).
\bibitem{Gid15} C. Gidney, \href{https://algassert.com/circuits/2015/06/12/Constructing-Large-Increment-Gates.html}{\textit{Constructing large increment gates}} (2015).
\bibitem{Bar95} A. Barenco, C. H. Bennett, R. Cleve, D. P. DiVincenzo, N. Margolus, P. Shor, T. Sleator, J. A. Smolin, and H. Weinfurter, \textit{Elementary gates for quantum computation}, Phys. Rev. A \textbf{52}, 3457 (1995).
\bibitem{Kuk25} P. Kuklinski, B. Rempfer, J. Elenewski, and K. Obenland, \textit{Efficient block-encodings require structure}, 	arXiv:2509.19667.
\bibitem{Cha19} S. Chakraborty, A. Gily\'{e}n, and S. Jeffery, \textit{The power of block-encoded matrix powers: improved regression techniques via faster Hamiltonian simulation}, Proc.\ 46th ICALP (2019).
\bibitem{Par20} J. B. Parker and I. Joseph, \textit{Quantum phase estimation for a class of generalized eigenvalue problems}, Phys. Rev. A \textbf{102}, 022422 (2020).
\bibitem{Lia22} J.-M. Liang, S.-Q. Shen, M. Li, and S.-M. Fei, \textit{Quantum algorithms for the generalized eigenvalue problem}, Quantum Inf. Process. \textbf{21}, 23 (2022).
\bibitem{Sha22} C. Shao and J.-P. Liu, \textit{Solving generalized eigenvalue problems by ordinary differential equations on a quantum computer}, Proc. R. Soc. A \textbf{478}, 20210797 (2022).
\bibitem{Raj26} G. Rajchel-Mieldzio\'c, S. Pli\'s, and E. Zak, \textit{Quantum algorithm for solving generalized eigenvalue problems with application to the Schr\"odinger equation}, arXiv:2506.13534.
\bibitem{Har09} A. W. Harrow, A. Hassidim, and S. Lloyd, \textit{Quantum algorithm for linear systems of equations}, Phys. Rev. Lett. \textbf{103}, 150502 (2009).
\bibitem{Bra23} C. Bravo-Prieto, R. LaRose, M. Cerezo, Y. Subasi, L. Cincio, and P. J. Coles, \textit{Variational quantum linear solver}, Quantum \textbf{7}, 1188 (2023).
\bibitem{Tra23} C. J. Trahan, M. Loveland, N. Davis, and E. Ellison, \textit{A variational quantum linear solver application to discrete finite-element methods}, Entropy \textbf{25}, 580 (2023).
\bibitem{Mor24} M. E. S. Morales, L. Pira, P. Schleich, K. Koor, P. C. S. Costa, D. An, A. Aspuru-Guzik, L. Lin, P. Rebentrost, and D. W. Berry, \textit{Quantum Linear System Solvers: A survey of algorithms and applications}, arXiv:2411.02522.
\bibitem{Wu06} C. Wu, B. A. Bernevig, and S.-C. Zhang, \textit{Helical liquid and the edge of quantum spin Hall systems}, Phys. Rev. Lett. \textbf{96}, 106401 (2006).
\bibitem{Zak24} V. A. Zakharov, J. Tworzyd{\l}o, C. W. J. Beenakker, and M. J. Pacholski, \textit{Helical Luttinger liquid on a space-time lattice}, Phys. Rev. Lett. \textbf{133}, 116501 (2024).
\bibitem{Wan23} Z. Wang, F. Assaad, and M. Ulybyshev, \textit{Validity of \textsc{slac} fermions for the $(1+1)$-dimensional helical Luttinger liquid}, Phys. Rev. B \textbf{108}, 045105 (2023).
\end{thebibliography}
\end{document}